# Ultrafast dynamics of coherent phonons and phonon-polaritons in lithium niobate crystals

Aizitiaili Abulikemu, Takumi Fukuda, and Muneaki Hase

*Department of Applied Physics, Faculty of Pure and Applied Sciences, University of Tsukuba, 1-1-1 Tennodai, Tsukuba 305-8573, Japan*

This study investigates the ultrafast dynamics of coherent phonons and phonon-polaritons in lithium niobate ($LiNbO_3$) crystals using reflective pump–probe spectroscopy with 25-fs time resolution. In addition to several coherent optical phonon modes, the electro-optic sampling measurements explored the coexistence of phonon-polariton *E*-modes near 3.8 and 14.9 THz, which agrees considerably with the theoretical phonon-polariton dispersion curves. We also discovered that a time lag ($\Delta t \approx 0.2$ ps) between the coherent optical phonon and phonon-polariton originates from the coupling time for propagating phonon-polariton. Our findings provide a possible application of $LiNbO_3$ for ultrafast electro-optic phonon modulators with bandwidths greater than 10 THz.

Ferroelectric crystals are superior materials in many respects, and lithium niobate (LiNbO$_3$) inherits most of their essential properties, such as excellent electro-optic (EO)[1] and photorefractive effects.[2] These features enable LiNbO$_3$ crystals to have various practical applications, including terahertz (THz) generators[3] and optical modulators.[4] Furthermore, LiNbO$_3$ is also one of the model systems for the generation and observation of a phonon-polariton; a quasiparticle generated by strong coupling between a photon and a transverse optical (TO) phonon. However, detailed exploration of ultrafast dynamics of phonon-polaritons has remained elusive, although it is essential for further extension of current optical and photonic applications of LiNbO$_3$ crystals. The generation and detection of coherent phonon-polaritons in LiNbO$_3$ crystals have been studied extensively using impulsive stimulated Raman scattering,[5] THz time-domain spectroscopy,[6] and THz emission spectroscopy.[7] However, these detection techniques exhibit numerous interference effects and have relatively low sensitivity and precision. Therefore, developing a more advanced approach is strongly required to explore coherent phonon-polariton dynamics in LiNbO$_3$ crystals. In contrast, time-resolved pump–probe spectroscopy is a useful technique for studying ultrafast coherent phonon dynamics with high sensitivity and high time resolution,[8,9] even for two-dimensional advanced materials.[10]

Some pioneering studies relating coherent phonon-polariton spectra in ferroelectric materials of the LiNbO$_3$ family with different structural defects and impurities have been reported in the past few decades.[11–13] However, most of them used transmission-type pump–probe ($\Delta T/T$) techniques with considerably long-pulse lasers (~100 fs or much longer). Such long laser pulses are insufficient to impulsively excite all phonon-polariton modes in LiNbO$_3$ crystals.[14,15] Compared with the transmission-type pump–probe measurements, the reflection-type pump–probe measurements with and without EO sampling can provide more precise measurements due to less scattering and short optical penetration, which help to gain information from the surface region of the sample, where backward propagated phonon-polaritons can exist.[16] In addition, the pump-probe EO sampling ($\Delta R_{eo}/R$) technique is a powerful tool for exploiting an anisotropic refractive index change associated with longitudinal optical (LO) phonon response,[17] although this technique has not been applied for ultrafast dynamics in LiNbO$_3$.

In this study, we investigated ultrafast dynamics of coherent phonon and phonon-polaritons simultaneously in LiNbO$_3$ crystals using the reflection-type pump-probe spectroscopy with and without EO sampling. We examined the fundamental physics behind the coexistence of coherent phonons and phonon-polaritons and the considerable time-lag between them.

A square slice of *x*-cut (5.0 × 4.0 × 0.2 mm$^3$) undoped LiNbO$_3$ single crystal was used in this study because it has shown significant potential as a medium for LiNbO$_3$-based EO modulators.[4] LiNbO$_3$ belongs to the ferroelectric R3c space group with $4A_1 \oplus 5A_2 \oplus 9E$ optical modes at room temperature, where the $A_1$- and $E$-symmetric TO and LO modes being simultaneously infrared (IR) and Raman active.[18] The measurements were performed using a homebuilt pump-probe system with reflection geometry. A mode-locked Ti: sapphire oscillator with a center wavelength of 820 nm (1.51 eV), a repetition rate of 80 MHz, and a pulse duration of ≤25 fs was used as a light source. The *s*-polarized pump (diameter of ≈19 μm) and the *p*-polarized probe (diameter of ≈15 μm) beams were co-focused onto the sample surface using an off-axis parabolic mirror, assuring nearly homogeneous excitation. On the other hand, the polarization of the pump beam was oriented along with the ordinary (y-axis) direction. This ccorresponds to the Porto notation in Raman scattering as $x(yz)\bar{x}$ and would allow us to observe the *E*-mode phonon-polaritons.[19] The optical fluence of the pump beam was fixed at $I_{ex} \approx 1.07$ mJ/cm$^2$, much lower than the damage threshold of LiNbO$_3$ crystals.[20] A balanced Si-PIN photodetector was used to selectively detect only the reflected probe beam (the transient reflectivity change Δ*R*/*R*). We inserted a polarizing beam splitter in front of the photodetectors for the EO sampling measurement. The polarization of the probe beam was controlled to 45° for the crystal *c*-axis [see Fig. 1(a) and (b)]. Meanwhile, Fig. 1(c) illustrates a schematic depiction of the current system's wave vector conservation of phonon-polaritons. The pump and probe beams are spatially overlapped in the LiNbO$_3$ crystal, and the pump wave vector is always assumed parallel to the *z*-direction. All measurements were performed in air at room temperature (295 K) conditions.

Figure 2 shows the results of the pump–probe reflectivity measurements from the *x*-cut LiNbO$_3$ crystal (a) without and (b) with EO sampling techniques. As the inset of Fig. 2(a) and (b) shows, we observed relatively strong coherent phonon oscillations between 0- and 1-ps time delays. The reflective signals significantly increase following the arrival of the pump pulse ($t = 0$ ps) on the surface of LiNbO$_3$ crystals due to a nonlinear optical response. Subsequently, we observed a sharp negative peak, followed by coherent phonon signals. The time evolution of the reflectivity change after the negative signal shows a double exponential decay background, and the signals of coherent phonon oscillations were obtained by subtracting the transient nonlinear signals and double exponential decay background. The Fourier transform (FT) spectra of the non-EO and EO sampling measurements were obtained from the coherent oscillatory signals. The FT spectra [Fig. 2(a)] shows a relatively broadband peak centered at 2.1 THz. To study the origin of the broadband peak, multiple-peak fits were employed using the Lorenz function, and the best fit was obtained for the three peaks at 1.34, 2.07, and 3.10 THz. As observed in previous experimental studies,[13,21] this broadband peak can be assigned to the symmetric $A_1$-mode phonon-polariton in LiNbO$_3$.

To investigate the anisotropic reflectivity changes of $x$-cut LiNbO$_3$ crystals, we performed EO sampling ($\Delta R_{eo}/R$) measurements because it could help us to observe anisotropic $E$-TO modes (by inducing a birefringent change of the refractive index), $A_1$-LO modes, and phonon-polariton modes simultaneously. As presented in Fig. 2(b), we emphasize the observed large enhancement of the intensity and coexistence of coherent TO and LO phonons and phonon-polaritons in the EO sampling measurements. The peak positions are located at 3.8, 7.2, 8.3, 10.1, and 14.9 THz. The intensity of the low-frequency peak (at 3.8 THz) is much stronger than that of the other peaks, with a narrower full width half maximum (FWHM) of $\Delta \nu \approx 0.3$ THz. The three signals near 7.2, 8.3, and 10.1 THz can be assigned to the coherent $2E$(TO), $1A_1$(LO), and $2A_1$(LO) modes, respectively. These observations are consistent with the results from transmission-type pump–probe measurement [15] and Raman scattering spectra[18, 22, 23] of LiNbO$_3$ single crystals.

To examine the possible assignment of the low-frequency (3.8 THz) and high-frequency (14.9 THz) modes as phonon-polariton modes, the phase-matching conditions of the $E$-mode phonon-polariton are considered, and the dispersion curve of the phonon-polaritons in LiNbO$_3$ are calculated for comparison based on the equation with no polariton damping[7, 14, 24]:

$$\epsilon(\Omega, k) = \frac{c^2 k^2}{\Omega^2} = \sum_{i=1}^{3} \frac{\epsilon_0 \Omega_{TOi}^2 - \epsilon_\infty \Omega^2}{\Omega_{TOi}^2 - \Omega^2}, \quad (1)$$

Where $\Omega$ is the frequency, $k$ is the wavevector, $c$ is the speed of light, $\Omega_{TOi}$ represents the frequency of the $i$-th TO phonon, $\epsilon_0$ (= 20.6) and $\epsilon_\infty$ (= 4.6) are the dielectric constants for low- and high-frequency limits, respectively.[18,24] We can obtain the dispersion of the phonon-polariton, *i.e.*, $\Omega_\pm$ by solving Eq. (1) for $\Omega$.

As shown in Fig. 3, the solid lines denote the dispersion calculated using Eq. (1), where the lowest three TO modes, TO$_1$ = 1.3 THz, TO$_2$ = 2.4 THz, and TO$_3$ = 3.4 THz, were included for the $A_1$-mode, and TO$_1$ = 4.5 THz, TO$_2$ = 7.1 THz, and TO$_3$ = 7.9 THz, were included for the $E$-mode. The phase-matching conditions for the forward (backward) propagating phonon-polariton wave vectors $k_+$ ($k_-$) can be calculated using the following equations[14,25]:

$$\Omega_+ = \frac{ck_+ + (n_e - n_o)f}{n_e}, \quad (2)$$

$$\Omega_- = \frac{ck_- + (n_o - n_e)f}{n_o}. \quad (3)$$

Here, $n_o$ (=2.26) and $n_e$ (=2.18) are the ordinary and extraordinary refractive indices of the sample, respectively[25]; $f$ is the frequency of the laser ($\approx$366 THz); and $\Omega_+$ ($\Omega_-$) is the frequency of the forward (backward) propagating phonon-polaritons. Eqs. (2) and (3) determine the phase-matching conditions for the phonon-polariton wave excited by the pump pulse, and they do not depend on the probing geometry, i.e., transmission or reflection. In our reflection-type pump–probe configuration, however, the backward propagating phonon-polaritons are preferably observed, as described below.

As displayed in Fig. 2(a), the non-EO sampling measurements exhibited a broader spectrum of the $A_1$ phonon-polariton mode originating from the overlapping of the three branches, including 1.34, 2.07, and 3.1 THz. These data points are in the range of the left-hand side ($\Omega_-$) of the phase-matching dispersion curve as shown in Fig. 3(a). Additionally, it can be confirmed that the data point of the high-frequency phonon-polariton mode at 14.9 THz is closer to the negative side of the phase-matching lines along the dispersion curve. We calculated the effective wave vectors ($k_+$ and $k_-$) for the observed phonon-polariton modes using Eqs. (2) and (3). For the 3.8-THz peak, the corresponding wave vector is $k_+ = 1238.4$ cm$^{-1}$, whereas the corresponding wave vector for the high-frequency broadband peak at 14.9 THz is $k_- = 73$ cm$^{-1}$. Thus, we assign both the intense low-frequency peak (at 3.8 THz) and the high-frequency broadband peak (at 14.9 THz) to originate from the *E*-mode phonon-polaritons, as shown in Fig. 3(b). The forward propagating wave vectors obtained in this study agree with the values reported by different research groups, based on the aforementioned calculation of the phase-matching conditions.[25,26] These results were induced via backward propagating phase-matching $\Omega_-$ because we employed reflection geometry, where backward propagating polariton would be better observed. In contrast, only the lowest *E*-phonon-polariton (3.8 THz) is closer to the $\Omega_+$ lines in the dispersion curves. This result may imply that the 3.8-THz mode originated from the forward propagating polariton.

To further investigate the dynamics of the low-frequency phonon-polariton mode, the time domain coherent oscillations of low-frequency *E*-mode (3.8 THz) phonon-polariton and $1A_1$(LO) mode (8.3 THz) were analyzed based on the short window FT (SWFT) using the following equation[27,28]

$$I(\omega, \tau) = \int_0^\infty dt \frac{\Delta R_{eo}(t)}{R} g(t-\tau) e^{-i\omega t}. \qquad (4)$$

Here, $\tau$ is the translation time and $I(\omega, \tau)$ is a complex-valued Fourier coefficient. The window function $g(t-\tau)$ is proportional to $e^{\frac{-(t-\tau)^2}{\sigma^2}}$ with $\sigma = \frac{W}{2\sqrt{\log 2}}$, where $W$ is the FWHM of the window width.[27]

In the SWFT, we employed the Gaussian window function with $W = 500$ fs window width and for the translation time step of 100 fs to make the signal smoothly decay in time and frequency. The SWFT spectra can be obtained using the FT of the series of the windowed oscillatory parts at each time delay. Figure 4(a) demonstrates the SWFT spectra of the coherent vibrations obtained in the time–frequency space. The overall signal intensity decreases as the time delay increases. Meanwhile, we still observed several characteristic peaks, including *E*-mode phonon-polariton and $1A_1$(LO)-mode, with some differences in the peak intensity. At $t = 0.0$ ps, the SWFT intensity of the $1A_1$(LO)-mode is significantly greater than that of the *E*-mode phonon-polariton, but it decreases with increasing time frame, suggesting that the *E*-mode phonon-polariton dominates over the $1A_1$(LO)-mode when the time frame is larger than 0.7 ps.

Figure 4(b) shows the time evolutions of the peak intensity for the phonon-polariton *E*-mode and $1A_1$(LO)-mode. The rise time of the amplitude of the *E*-mode phonon-polariton is slower than that of the $1A_1$(LO)-mode, whereas the *E*-mode phonon-polariton exhibits a longer relaxation time than the $1A_1$(LO)-mode. Interestingly, we observed a significant time lag ($\Delta t \approx 0.2 \pm 0.03$ ps) between the low-frequency *E*-mode phonon-polariton mode (3.8 THz) and the $1A_1$(LO)-mode (8.3 THz). To check the validity of the observed time lag, we examined various widths of the window in the analysis, such as 300, 400, 600, and 700 fs. As the results, the time-lag between the *E* and $A_1$ modes was always constant (data not shown), although the SWFT intensity slightly increases with increasing window width from 300 to 700 fs. Thus, we argue that the observed time-lag is independent on the parameters used in the SWFT analyses. In addition, we examined a possibility of over subtraction of transient nonlinear signals and double exponential decay background, and the results indicate the subtraction does not affect the observed time-lag. The obtained time-lag could be due to the time delay between the localized zone-center $A_1$(LO) mode and forward propagating phonon-polaritons in the $LiNbO_3$ crystal because the probe beam needs additional time to see the forward propagating wave.[14] Note that the build-up time of 100 fs for the formation of plasmon-phonon coupling in semiconductors was observed.[29] In general, the coupling time for the phonon-polariton mode is governed by $T = 2\pi/\Omega_{TO}$. In the present case, we obtain $T \approx 230$ fs, which is close to the time-lag (200 fs). However, further experimental and theoretical studies are required to explore the dynamics of the observed time-lag.

In conclusion, we have explored the dynamical properties of coherent phonons and phonon-polaritons in $LiNbO_3$ ferroelectric crystals using the pump–probe reflectance spectroscopy with and without EO sampling. We compared the frequencies of the observed coherent phonons and phonon-polaritons with the dispersion relations and observed the coexistence of the LO modes and TO phonon oscillation and low- and high-frequency *E*-mode phonon-polaritons at 3.8 and 14.9 THz, respectively. Furthermore, a time lag ($\Delta t \approx$ 0.2 ± 0.03 ps) was obtained between $1A_1$(LO) mode (at 8.3 THz) and low-frequency *E*-mode phonon-polariton (at 3.8 THz), which is considered to originate from the coupling time of the forward propagating phonon-polaritons in the *x*-cut $LiNbO_3$ crystals. Our demonstration paves the way for using this approach to other ferroelectric crystals such as $BaTiO_3$, and $PbTiO_3$. Moreover, we expect that these results will benefit the development of $LiNbO_3$-based EO devices, including Mach–Zehnder-type EO modulators,[4] optical single-sideband modulators,[30] and EO frequency comb,[31] which are essential for realizing information conversion between electronic and optical domains in data communication systems.

**Acknowledgments**

This study was supported by Core Research for Evolutional Science and Technology program of the Japan Science and Technology (Grant Number: JPMJCR1875). T. Fukuda is supported by JSPS Research Fellowships for Young Scientists and JSPS KAKENHI (Grant Number: 21J20332). We thank Dr. Jessica Afalla for useful discussions.

## Figure Captions

**Fig. 1.** (a) Simplified scheme of the reflective pump-probe spectroscopy based on electro-optic sampling. (b) Schematic diagram of *x*-cut $LiNbO_3$ crystal structure. (c) Schematic illustration of wave vector conservation.

**Fig. 2.** The Fourier transformed (FT) spectra of the coherent oscillatory signals were detected using (a) non-EO and (b) EO sampling geometry. The insets represent time-resolved reflectivity signals (the solid lines represent the raw data, and the dotted lines represent an exponential curve fit).

**Fig. 3.** The dispersion relations of phonon-polaritons in $LiNbO_3$ crystals for (a) the $A_1$-mode and (b) the *E*-mode. The double dashed lines represent the phase-matching conditions, $\Omega_+$ ($\Omega_-$), including the possible errors ($\pm 0.005$) of $n_o$ and $n_e$. The blue horizontal lines represent the bare frequency of $TO_1 = 1.3$ THz, $TO_2 = 2.4$ THz, and $TO_3 = 3.4$ THz for the $A_1$-mode and $TO_1 = 4.5$ THz, $TO_2 = 7.1$ THz, and $TO_3 = 7.9$ THz for the *E*-mode. The closed circles represent the phonon-polariton peak frequencies observed in Fig. 2.

**Fig. 4.** (a) The short window Fourier transform (SWFT) spectra of coherent vibrations obtained in time–frequency space. (b) The intensity of SWFT spectra of the phonon-polariton *E*-mode (3.8 THz) and $1A_1$(LO)-mode as the function of the time delay.

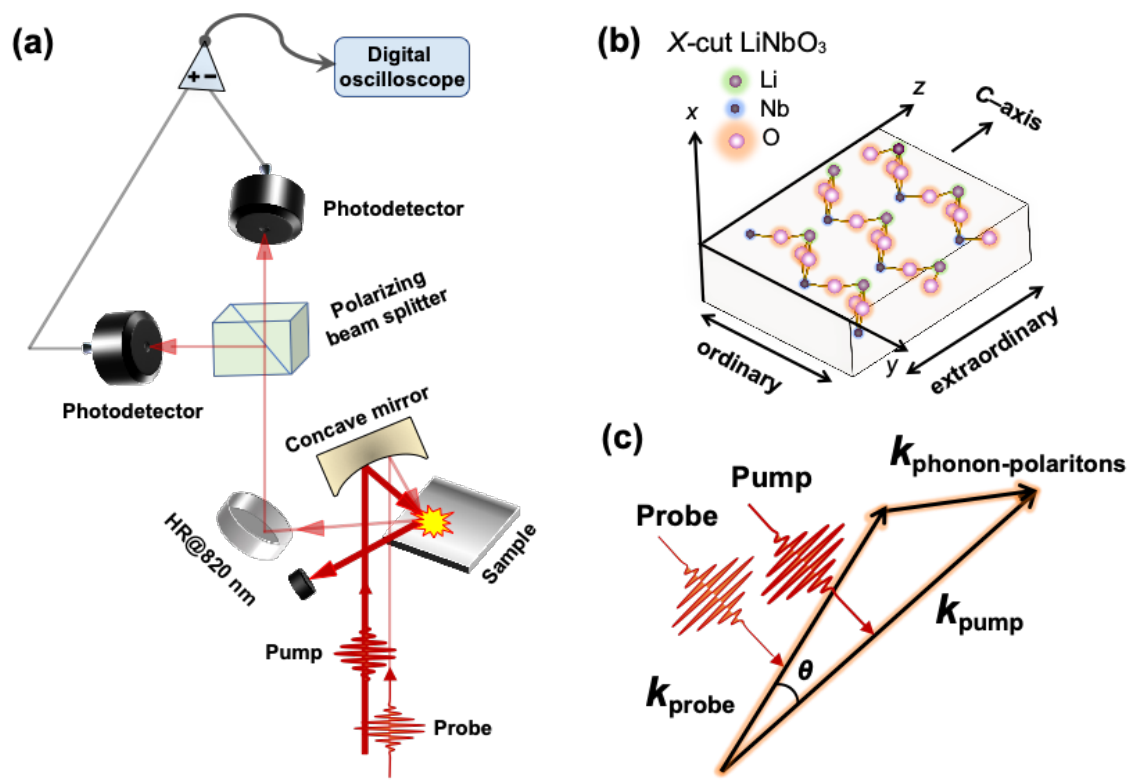

**Fig. 1.** A. Abulikemu *et al.*

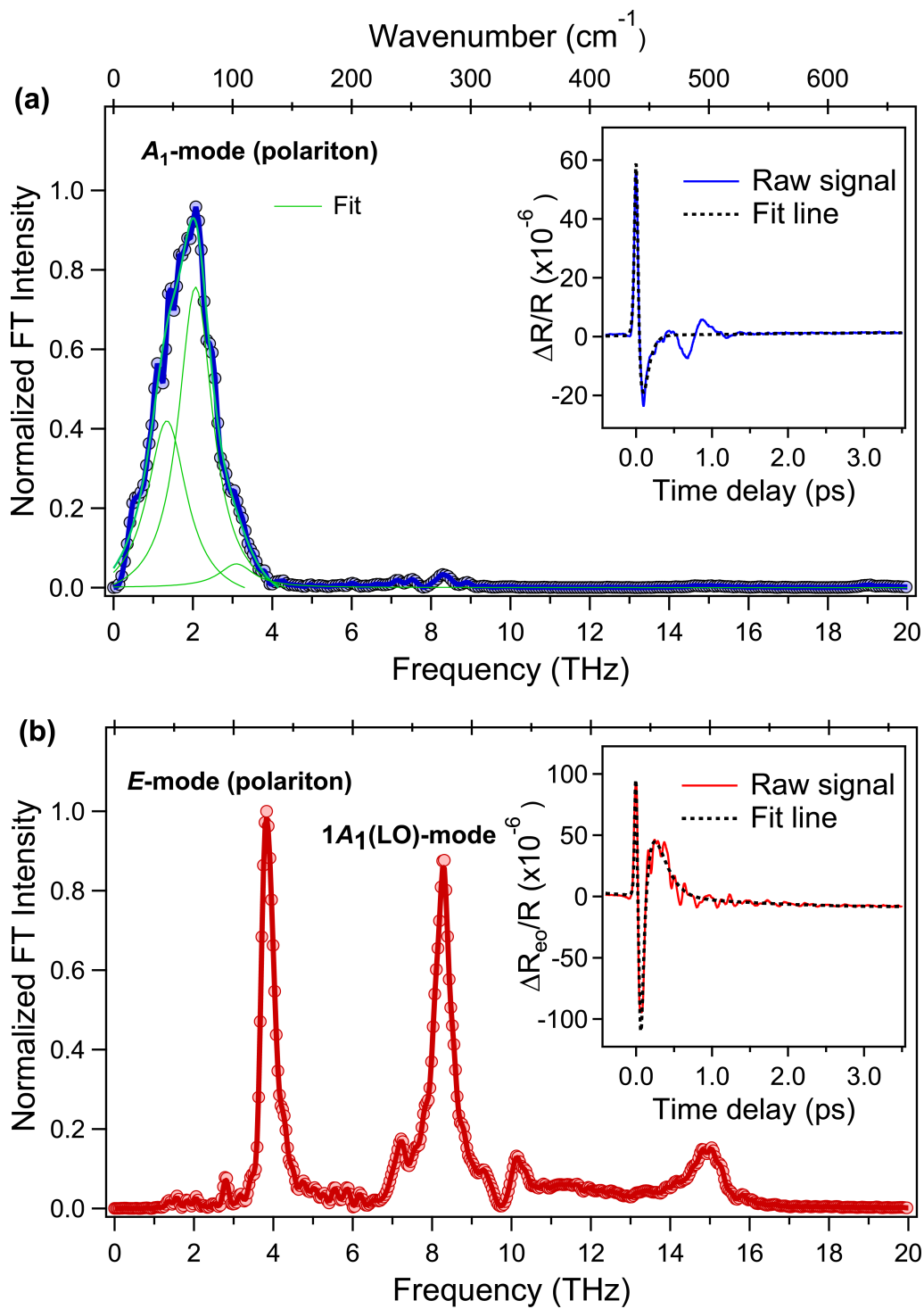

**Fig. 2.** A. Abulikemu *et al.*

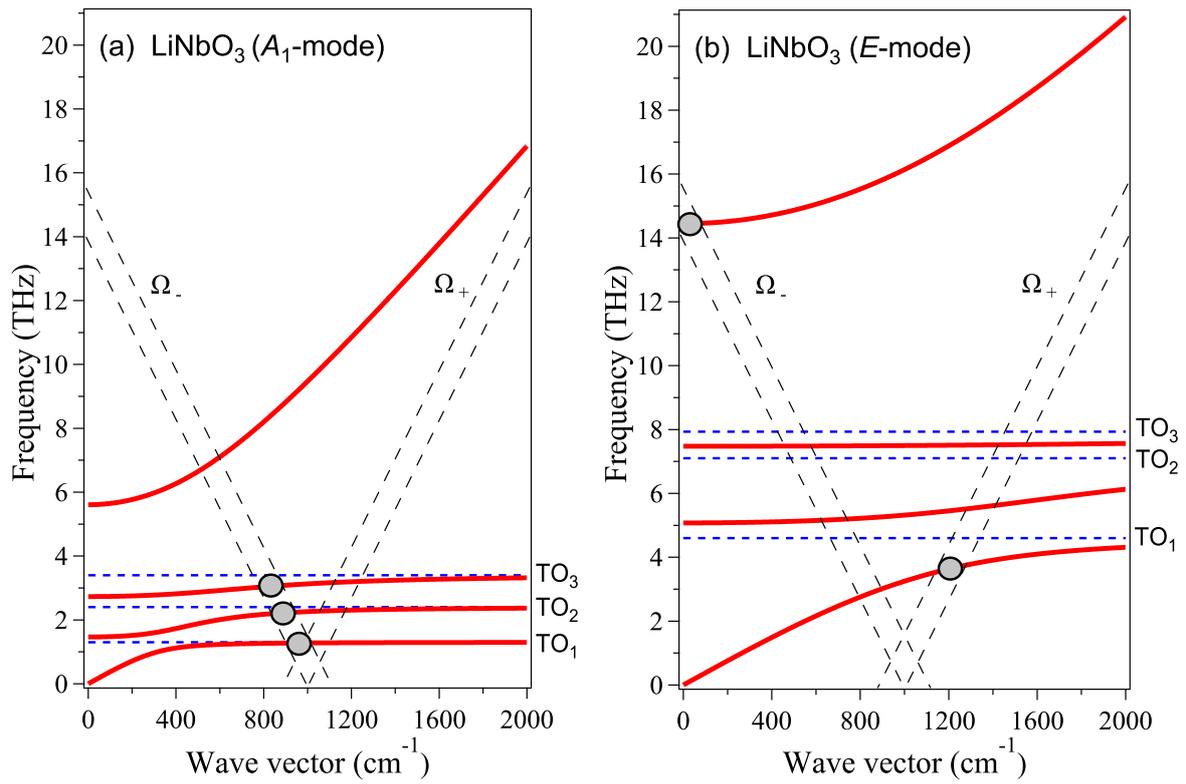

**Fig. 3.** A. Abulikemu *et al.*

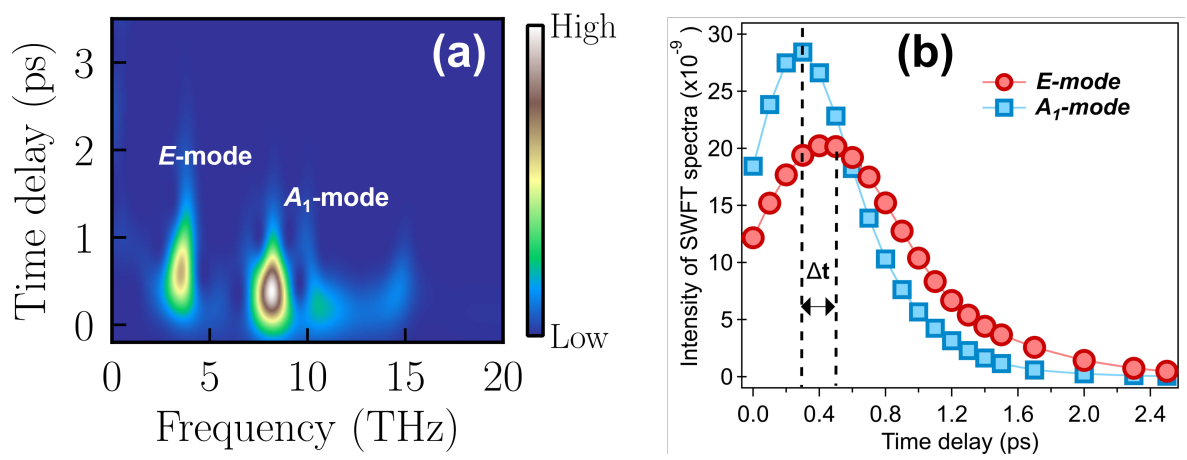

**Fig. 4.** A. Abulikemu *et al.*